\documentclass[aps,pre,twocolumn,superscriptaddress]{revtex4-1}
\usepackage{graphicx}
\usepackage{dcolumn}
\usepackage{bm}
\renewcommand\vec[1]{\boldsymbol{#1}}
\usepackage{amsmath}
\usepackage{amssymb}
\usepackage{color}
\begin{document}


\title{Phase transitions and ordering structures
of a model of chiral helimagnet\\
in three dimensions 
}


\author{Yoshihiko Nishikawa}
\email{nishikawa@huku.c.u-tokyo.ac.jp}
\affiliation{Department of Basic Science, University of Tokyo\\
3-8-1 Komaba, Meguro, Tokyo 153-8902, Japan}
\author{Koji Hukushima}
\email{hukusima@phys.c.u-tokyo.ac.jp}
\affiliation{Department of Basic Science, University of Tokyo\\
3-8-1 Komaba, Meguro, Tokyo 153-8902, Japan}
\affiliation{Center for Materials Research by Information Integration,
National Institute for Materials Science, 1-2-1 Sengen, Tsukuba,
Ibaraki 305-0047, Japan}


\date{\today}

\begin{abstract}
Phase transitions in a classical Heisenberg spin model 
of a chiral helimagnet
with the Dzyaloshinskii--Moriya (DM) interaction
in three dimensions
are numerically studied.
By using the event-chain Monte Carlo algorithm
recently 
developed for particle and continuous spin systems,
we perform equilibrium Monte Carlo simulations 
for large systems up to about $10^6$ spins. 
Without magnetic fields,
the system undergoes a continuous phase transition
with critical exponents of the three-dimensional \textit{XY} model,
and a uniaxial periodic helical structure emerges in the low temperature region.
In the presence of a magnetic field perpendicular to the axis of the helical structure,
it is found that 
there exists a critical point 
on the temperature and magnetic-field  phase diagram
and that
above the critical point the system exhibits a phase transition with
strong divergence of the specific heat and the uniform magnetic susceptibility. 
\end{abstract}


\maketitle


\section{Introduction}

Frustration and competition 
between interactions and/or fields
often induce
complicated spin structures
into magnetic materials
such as
spin ice, magnetic skyrmion,
and spin liquid. 
Phase transitions and phase diagrams in magnetic materials 
driven by various interactions and fields
have been extensively studied in condensed matter physics and also
statistical physics. Among them, 
chiral magnets
such as MnSi
have recently attracted great interests
to experimental and theoretical studies
not only for its fundamental properties
but also for applications 
\cite{KIY, CSL_CHM_ex1, CSL_CHM_ex2, CSL_CHM_ex3, ZMJDGZPTZ, IEDzyalo64, IEDzyalo64_2, IEDzyalo65,  CSL_CHM2}.
Chiral helimagnet is a magnetic system in which 
a uniaxial helical structure emerges in the low temperature region.
The helical structure is induced 
by the Dzyaloshinskii--Moriya (DM) interaction \cite{DM1, DM2}
which is an antisymmetric interaction breaking a chiral symmetry, and thus,
the two same helical structures with different winding directions
do not degenerate.
By a variational analysis 
of a one-dimensional continuum model \cite{IEDzyalo64, IEDzyalo64_2, IEDzyalo65, CSL_CHM2},
it is revealed theoretically that
a chiral magnetic soliton lattice (CSL) is formed 
with a finite magnetic field 
perpendicular to the axis of the helical structure,
and a continuous phase transition to forced ferromagnetic phase occurs
with increasing the magnetic field. 
A mean-field analysis shows
that a phase transition into the CSL phase 
occurs at a finite temperature under the
magnetic field \cite{CHM_MFT}.

While recent experiments \cite{CSL_CHM_ex1, CSL_CHM_ex2}
have reported the existence of the CSL state at finite temperatures in three dimensions,
finite-dimensional effects beyond the mean-field theory
on the nature of the finite-temperature phase transitions
of the system 
are still less clear.
In the absence of magnetic fields,
renormalization-group approaches \cite{RG_CHM1, RG_CHM2}
predict that the system undergoes
a continuous phase transition
with critical exponents of the ferromagnetic \textit{XY} model.
Another theoretical study \cite{DM-XY}
also indicates that 
the system belongs to the same universality class of the ferromagnetic \textit{XY} model.
On the other hand, 
with the magnetic field 
perpendicular to the axis of the helical structure,
the system no longer has any continuous symmetry in the spin space. 
Therefore, the nature of a possible phase transition in three dimensions
is nontrivial 
and possibly different from the three-dimensional \textit{XY} model.

In this paper, 
we study a three-dimensional classical Heisenberg spin model of a chiral helimagnet 
by equilibrium Monte Carlo simulations. 
We especially focus on its phase transitions 
and ordering structures in the low temperature region
with and without the magnetic field. 
Because of 
the competition among 
the DM interaction, 
the symmetric exchange interaction, 
and the magnetic field, 
complicated ordering structures
emerge in the low temperature region.
In particular,
there are many CSL states 
with different numbers of chiral solitons 
which are separated with each other by large energy barrier. 
Hence, a transition between the different CSL states hardly occurs by
means of conventional Monte Carlo algorithms 
such as the Metropolis and the heat-bath algorithm.  
In order to reduce the difficulty of the slow relaxation, 
we use the event-chain Monte Carlo algorithm
\cite{originalECMC, generalECMC1, generalECMC2, ECMC_XY, ECMC_Heisenberg}
which is a recently proposed rejection-free and 
efficient algorithm
for equilibrium simulations.
This algorithm
enables us to 
equilibrate quite large systems
with more than $10^6$ spins
so as to avoid suffering from 
its strong finite-size effects
particularly 
in the presence of the magnetic field.  

This paper is organized as follows.
In Section~\ref{sec:Model}
we define a classical Heisenberg spin model 
of a chiral helimagnet
and various physical quantities.
The details of the event-chain Monte Carlo algorithm
are presented in Section~\ref{sec:ECMC}. 
In Section~\ref{sec:result},
results of our Monte Carlo simulations are shown, 
and properties of phase transitions
and ordering structures
of the system with and without a magnetic field
are discussed.
In Section~\ref{sec:discussion}
we discuss a possible phase diagram
and summarize our results.


\section{Model and physical quantities}
\label{sec:Model}
In this paper, we study a classical Heisenberg model of a chiral helimagnet 
in a three-dimensional simple cuboidal lattice.
The system is defined by the Hamiltonian
\begin{eqnarray}
\label{eq:Hamiltonian}
H\left(\left\{ \vec{S}_i\right\}\right) &=& -J\sum_{\left\langle i,j \right\rangle} \vec{S}_i \cdot \vec{S}_j
- \vec{D}  \cdot \sum_i \left( \vec{S}_i \times \vec{S}_{i+\hat{y}}\right) \nonumber\\
&&\hspace{3cm}- \vec{h} \cdot \sum_i \vec{S}_i,
\end{eqnarray}
where $\vec{S}_i$ is a unit vector with three components,
$J$ is a positive coupling constant, 
$\vec{D} = D\hat{y}$ is the DM vector,
and $\vec{h} = h \hat{z}$ is a magnetic field perpendicular to the DM vector $\vec{D}$.
The summation in the first term runs over all the neighboring pairs of sites,
and the other summations run over all the sites.
The lattice on which the system is defined
is a cuboid where the linear size of $y$ direction is 
$\alpha$ times as long as $x$ and $z$ directions.
The linear size of $x$ and $z$ directions of the lattice is denoted by $L$
and the total number of sites is $N = \alpha L^3$.
We set $\alpha = 8$ in the following of this paper.
Periodic boundary conditions 
are imposed on $x$ and $z$ directions
and a free boundary condition on $y$ direction.

The second term in the Hamiltonian~(\ref{eq:Hamiltonian})
represents the Dzyaloshinskii--Moriya interaction \cite{DM1, DM2}
which induces a helical spin structure. 
In the ground state of the system without magnetic fields,
all spins in each $x$-$z$ plane align 
ferromagnetically
and
the spins in each plane make a canted 
angle $\theta=\arctan(D/J)$  with respect to its nearest neighbor plane
along the DM vector. 
The wave vector $\vec{q}_\mathrm{chiral}$ corresponding to the helical structure in the ground state
is determined by $D / J$ via
\begin{equation}
\vec{q}_\mathrm{chiral} = \arctan \left( \frac{D}{J} \right) \hat{y}.
\end{equation}
At a finite temperature, 
the system undergoes a phase transition
from a paramagnetic phase
to a chiral helimagnetic phase as temperature decreases.
Following the work by Calvo \cite{DM-XY},
the system without magnetic fields
can be exactly mapped onto another system
defined by the Hamiltonian
\begin{equation}
\label{eq:transfHam}
H^\prime\left(\left\{ \vec{S}_i\right\}\right) 
= -J \sum_{\left\langle i,j \right\rangle_{\perp}} \vec{S}_i \cdot \vec{S}_j
- \sum_i \vec{S}_i \cdot C \vec{S}_{i+\hat y},
\end{equation}
where 
\begin{equation}
C = \left(
\begin{array}{ccc}
\sqrt{J^2 + D^2} & &\\
& J&\\
& & \sqrt{J^2 + D^2}
\end{array}\right),
\end{equation}
and the summation in the first term
runs over all the neighboring pairs of two sites
which are in the same $x$-$z$ plane.
This Hamiltonian~(\ref{eq:transfHam}) for a finite value of $D$ 
has the same symmetry with the \textit{XY} model,
and therefore,
the original system is expected 
to belong to the same universality class
of the three-dimensional ferromagnetic \textit{XY} model \cite{DM-XY}.

In the presence of the magnetic field $\vec{h}$
perpendicular to the DM vector,
the structure of the ground state is modulated
depending on $h = \left| \vec{h} \right|$.
For $0 < h < h_\mathrm{c}$, the CSL is formed \cite{CSL_CHM2},
and all spins are parallel to the magnetic field for
$h > h_\mathrm{c}$.
In the CSL state at zero temperature,
there are more than one local length scales such as
the distance between two chiral solitons and the length of one chiral soliton,
and hence, multiple wave vectors are expected to be
required to characterize the CSL structure.

For the chiral helimagnetic system,
we define the wave-vector-dependent magnetization
which captures the helical structure of the system
as
\begin{equation}
\vec{m}\left(\vec{q}\right) = \frac{1}{N}\sum_i \vec{S}_i \exp(\mathrm{i} \vec{q}\cdot \vec{r}_i),
\end{equation} 
where $\vec q$ is a three-component wave vector.
The wave-vector-dependent susceptibility associated with $\vec m \left( \vec q \right)$
is defined as
\begin{equation}
\chi\left(\vec{q}\right) 
= \beta N\left(\left\langle \left| \vec{m}\left(\vec{q} \right) \right|^2 \right\rangle 
- \left| \left\langle  \vec{m}\left(\vec{q}\right)\right\rangle \right|^2\right),
\end{equation}
where $\beta$ is an inverse temperature and the bracket $\left\langle \cdots \right\rangle$
denotes the thermal average.
Note that $\chi \left( \vec q\right)$ is proportional to a Fourier component
of the spin correlation function
\begin{equation}
C\left(\vec r \right) 
= \frac1N \sum_i \left(\left\langle \vec S_i \cdot \vec S_{i+\vec r} \right\rangle
- \left\langle \vec S_i\right\rangle \cdot \left\langle\vec S_{i+\vec r} \right\rangle
 \right).
\end{equation}
In particular, the susceptibility 
with a wave vector $\vec{q}$ parallel to 
the DM vector $\vec D$ is denoted as $\chi^{\parallel}\left( q \right)$,
where $q = \left| \vec q \right|$.
Although the ground state of the system
with no magnetic fields
is obviously characterized by
$\vec m \left( \vec q = \vec q_\mathrm{chiral} \right)$,
it is unclear that
which $\vec q$'s characterize the structure
at finite temperature with/without a magnetic field $\vec h \neq \vec 0$.
We thus calculate the wave-vector dependence 
of $\chi^\parallel \left( q \right)$, which yields 
the wave vectors $\vec q_0$ 
at which $\chi^\parallel \left( q_0 \right)$ gives a maximum value.
By using $\chi\left(\vec{q}\right)$,
the wave-vector-dependent finite-size correlation length
is defined as
\begin{equation}
\label{eq:def_corrlength}
\xi_L \left(\vec{q}\right) 
= \frac{1}{2\sin \left(\left| \vec{q}_\mathrm{min}\right| / 2 \right)}
\sqrt{\frac{\chi \left(\vec{q} \right)}{\chi \left(\vec{q} +\vec{q}_\mathrm{min}\right)} - 1},
\end{equation}
where $\vec{q}_\mathrm{min}$ is the minimum wave vector
parallel to $\vec{q}$.
Similarly to the susceptibility, 
the finite-size correlation length depending on 
a wave vector $\vec{q}$ parallel to $\vec D$ 
is defined as $\xi_{L}^{\parallel} \left(q \right)$,
where $\vec{q}_\mathrm{min}$ in Eq.~(\ref{eq:def_corrlength})
is set to $\vec{q}_\mathrm{min} = \left( 0, 2\pi / \alpha L, 0 \right)$.

We also define a distribution function of the energy density $e$ as
\begin{equation}
P\left( e \right) 
= \left\langle 
\delta\left( e - \frac{1}{N} H\left(\left\{ \vec S_i\right\}\right)\right) 
\right\rangle, 
\end{equation}
which is evaluated by Monte Carlo simulations. 
From the distribution, 
the specific heat $c$ is calculated. 
When the system exhibits 
a first-order phase transition, 
the distribution has a double-peak structure
at the transition temperature. 

We study the phase transitions of the system
with $D / J = 1$
by equilibrium Monte Carlo (MC) simulations
using the event-chain Monte Carlo (ECMC) algorithm
\cite{originalECMC, generalECMC1, generalECMC2, ECMC_XY, ECMC_Heisenberg}
combined with the heat-bath algorithm, the over-relaxation updates \cite{OR1,OR2} 
and the exchange Monte Carlo method (or parallel tempering) \cite{HukushimaNemoto}.
The details of the ECMC algorithm in our simulations are presented
in the next section.


\section{Event-chain Monte Carlo algorithm}
\label{sec:ECMC}
The ECMC algorithm
was originally developed for particle systems \cite{originalECMC, generalECMC1, generalECMC2},
and recently applied to continuous spin systems \cite{ECMC_Heisenberg, ECMC_XY}.
In every step of the algorithm,
only one particle (or spin) is moved,
and another interacting particle (or spin)
starts to move instead of rejecting a proposal.
Thus, a series of local updates called ``event chain'' is formed,
in which 
many particles (or spins) are updated in a cooperative manner.
This dynamics breaks the detailed balance condition,
but still satisfies the global balance condition.
For various systems, 
the ECMC algorithm outperforms conventional algorithms 
such as the Metropolis algorithm \cite{Metropolis}
and the heat-bath algorithm \cite{heatbath1, heatbath2}.
In particular, it is revealed 
that the algorithm reduces the value of the dynamical critical exponent $z$
of the three-dimensional ferromagnetic Heisenberg model 
to $z \simeq 1$
from the conventional value $z \simeq 2$ \cite{ECMC_Heisenberg}.
This reduction enables us to simulate 
systems with much larger degrees of freedom in equilibrium than those
attained with the conventional algorithms previously. 

In this algorithm, the state of the system 
is represented by $\left( \left\{\vec{S}_i \right\}, U\right)$,
where $\left\{\vec{S}_i \right\}$ is the spin configuration
and $U$ is a ``lifting parameter.''
The lifting parameter $U$
specifies
the current rotation site and the direction vector of the
rotation axis. 
Explicitly, the lifting parameter is given as an $N \times 3$ matrix of
the form  
$U = \vec{e}_j\vec{v}^\mathrm{T}$, 
where $\vec{e}_j$ is an $N$-dimensional unit vector 
with components $(\vec{e}_j)_k = \delta_{j,k}$ and 
$\vec{v}$ is a three-component unit vector. 
For concreteness, 
we assume that the Hamiltonian can be written
as a summation of interactions
\begin{equation}
\label{eq:HamDecomp}
H\left( \left\{ \vec{S}_i\right\} \right) 
= \frac12 \sum_{i, j} \sum_a E_{ij}^{(a)}\left(\vec{S}_i,\vec{S}_j \right)
+ \sum_{i,a} E_i^{(a)} \left( \vec S_i\right), 
\end{equation}
where the suffix ``$a$'' is the type of interaction. 
Note that any decompositions of the Hamiltonian
in the form of Eq.~(\ref{eq:HamDecomp})
are allowed in the following argument.
An elementary step of this algorithm is
to propose an infinitesimal rotation $\mathrm{d} \phi$ of the moving spin $\vec{S}_j$
around the axis $\vec{v}$,
and to accept the proposal with probability
of the factorized Metropolis filter \cite{generalECMC2}
\begin{eqnarray}
&&W_{U}\left( \mathrm{d}\phi \right)  \nonumber \\
&&= \prod_{\substack{k \in \partial j\\ a}} 
\exp\left( -\beta \max\left[ 
\left. \frac{\mathrm{d} \left(\Delta E_{jk}^{(a)}\left( \varphi=0 ;\vec{v}\right)\right) }{\mathrm{d} \varphi}\right., 0
\right] \mathrm{d}\phi \right)\nonumber \\
&&\times \prod_{a}
\exp\left( -\beta \max\left[ 
\left. \frac{\mathrm{d} \left(\Delta E_{j}^{(a)}\left( \varphi=0 ;\vec{v}\right)\right) }{\mathrm{d} \varphi}\right., 0
\right] \mathrm{d}\phi \right),\nonumber
\end{eqnarray}
where $\partial j$ means the set of sites interacting with $j$-th spin,
\begin{eqnarray}
\Delta E_{jk}^{(a)}\left( \varphi ; \vec{v}\right) &=& E_{jk}^{(a)}\left( R_{\vec{v}}\left( \varphi \right)\vec{S}_j, \vec{S}_k \right) - E_{jk}^{(a)}\left( \vec{S}_j, \vec{S}_k \right),\nonumber \\
\Delta E_{j}^{(a)}\left( \varphi ; \vec{v}\right) &=& E_{j}^{(a)}\left( R_{\vec{v}}\left( \varphi \right)\vec{S}_j \right) - E_{j}^{(a)}\left( \vec{S}_j \right),\nonumber
\end{eqnarray}
and $R_{\vec{v}}(\varphi)$ is a rotation matrix around $\vec{v}$ with an
angle $\varphi$. 
Thanks to the factorization,
whether the proposal is accepted
can be determined by each factor
independently, 
i.e., the proposal is accepted 
only if all the factorized potentials avoid the rejection. 
When the proposal is rejected by a factor 
with the potential $E_{jk}^{(a)}$ (or $E_{j}^{(a)}$), 
then a lifting event occurs 
and the lifting parameter is updated as $U \rightarrow L_{jk}^{(a)}U$
(or $U \rightarrow L_{j}^{(a)}U$),
where $L_{jk}^{(a)}$ (or $L_{j}^{(a)}$) is a lifting matrix.
The balance condition 
requires that 
$L_{jk}^{(a)}$ and $L_{j}^{(a)}$satisfy \cite{generalECMC1}
\begin{eqnarray}
\label{eq:liftingoperator1}
L_{jk}^{(a)}\vec{g}_{jk}^{(a)} &=& -\vec{g}_{jk}^{(a)},\\
\label{eq:liftingoperator2}
L_{j}^{(a)}\vec{g}_{j}^{(a)} &=& -\vec{g}_{j}^{(a)},
\end{eqnarray}
where
\begin{eqnarray}
\vec{g}_{jk}^{(a)}
&=&\frac{\mathrm{d}}{\mathrm{d}\varphi}
\left.\left( \Delta E_{jk}^{(a)}\left( \varphi; \vec v \right)\vec{e}_j
+ \Delta E_{kj}^{(a)}\left( \varphi; \vec v \right)\vec{e}_{k}\right)\right|_{\varphi=0}, \nonumber \\
\vec{g}_{j}^{(a)}
&=&\frac{\mathrm{d}}{\mathrm{d}\varphi}
\left.\left( \Delta E_{j}^{(a)}\left( \varphi; \vec v \right)\vec{e}_j
\right)\right|_{\varphi=0}, \nonumber
\end{eqnarray}
respectively. 
In general, 
$L_{jk}^{(a)}$ and $L_{j}^{(a)}$
which satisfy Eq.~(\ref{eq:liftingoperator1}) and
Eq.~(\ref{eq:liftingoperator2})
are rewritten 
by using an $N \times N$ regular matrix $A$  and the identity matrix
$\mathrm{I}$ as
\begin{eqnarray}
\label{eq:liftingop}
L_{jk}^{(a)} 
&=& \mathrm{I} - 
2\frac{A\vec{g}_{jk}^{(a)} \left(\vec{g}_{jk}^{(a)}\right)^{\mathrm{T}} }
{\vec{g}_{jk}^{(a)} \cdot A\vec{g}_{jk}^{(a)}}, \\
L_{j}^{(a)} 
&=& \mathrm{I} - 
2\frac{A\vec{g}_{j}^{(a)} \left(\vec{g}_{j}^{(a)}\right)^{\mathrm{T}} }
{\vec{g}_{j}^{(a)} \cdot A\vec{g}_{j}^{(a)}}.
\label{eq:liftingop2}
\end{eqnarray}
In principle, any matrix $A$ is available but a class of $A$ leading to
a simple lifting event is desired in practice. 
To make the algorithm into practice,
an event-driven approach \cite{event-driven} is adopted,
which allows to move the spins with a finite displacement. 

In the conventional ECMC algorithm for 
continuous spin systems only with isotropic interactions \cite{ECMC_Heisenberg,ECMC_XY}
and a magnetic field, 
the Hamiltonian is decomposed as
\begin{equation}
H_\mathrm{iso} \left( \left\{ \vec S_i \right\} \right) 
= \frac12 \sum_i \sum_{j \in \partial i} E_{ij} \left(  \vec S_i, \vec S_j \right)
+ \sum_i E_{i} \left( \vec S_i \right),
\end{equation}
where
\begin{eqnarray}
E_{ij} \left( \vec S_i, \vec S_j \right) &=& -J_{ij} \vec S_i \cdot \vec S_j, \\
E_{i} \left( \vec S_i \right) &=& -\vec h \cdot \vec S_i.
\end{eqnarray}
The isotropic interactions have a simple relation as
\begin{equation}
\label{eq:isotropic}
\left.\frac{\mathrm{d}}{\mathrm{d}\varphi}\Delta E_{jk}\left(\varphi; \vec v \right)\right|_{\varphi=0} 
= \left.-\frac{\mathrm{d}}{\mathrm{d} \varphi} \Delta E_{kj}\left(\varphi; \vec v \right)\right|_{\varphi=0}
\end{equation}
for all $j$, $k$ and $\vec v$.
This relation yields that 
by choosing 
the matrix $A$ in Eq.~(\ref{eq:liftingop}) and Eq.~(\ref{eq:liftingop2})
as the identity $\mathrm{I}$,  
the lifting matrices are determined as 
\begin{eqnarray}
\left( L_{jk} \right)_{p,q} &=& \delta_{p,q} - \delta_{j,p}\delta_{k,q} + \delta_{j,q}\delta_{k,p},\\
\left( L_{j} \right)_{p,q} &=& \delta_{p,q} \left( 1 - 2\delta_{j,p}\right), 
\end{eqnarray}
respectively. 
These lifting matrices
make the lifting parameter $U$
have one non-zero row,
and thus, only a single spin moves at any time.
However, 
for anisotropic interactions including the DM interaction,
Eq.~(\ref{eq:isotropic}) does not hold in general.
In these cases, 
$L_{jk}^{(a)}$ depends on the spin configuration,
and the updated lifting parameter $L_{jk}^{(a)} U$ 
has more than one non-zero rows,
meaning that multiple spins start to move after a lifting event.
Although we could implement another Monte Carlo algorithm
in which multiple spins move simultaneously \cite{generalECMC1, BPS},
we apply the ECMC algorithm 
only with the rotation axis $\vec{v} = \hat{y}$,
where Eq.~(\ref{eq:isotropic}) holds for the DM interaction
and thus the single spin update is still kept.
Instead, the ergodicity condition is not satisfied by the ECMC algorithm
only with a single rotation axis.
In order to recover the ergodicity condition in the Markov chain, 
the over-relaxation 
and the heat-bath algorithms 
are combined with this ECMC algorithm.
The ECMC 
algorithm
enables us to sample different structures of the system efficiently by 
inducing 
cooperative spin updates
of
the same $x$-$z$ plane
in each event chain. 

\begin{figure}[t]
\includegraphics[width=0.8\linewidth]{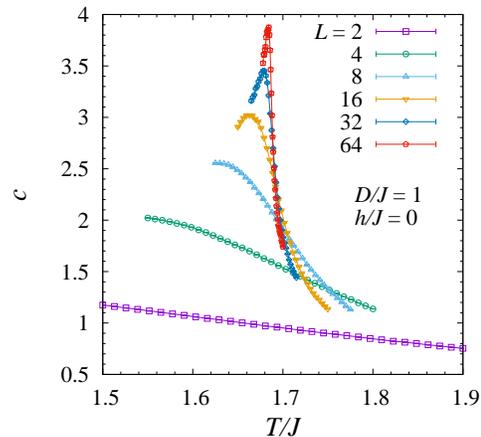}
\caption{(Color online) 
Temperature dependence of the specific heat $c$
of the chiral helimagnetic model in three dimensions without magnetic fields. 
 }
\label{fig:static_h0}
\end{figure}

\begin{figure}[!t]
\includegraphics[width=1.0\linewidth]{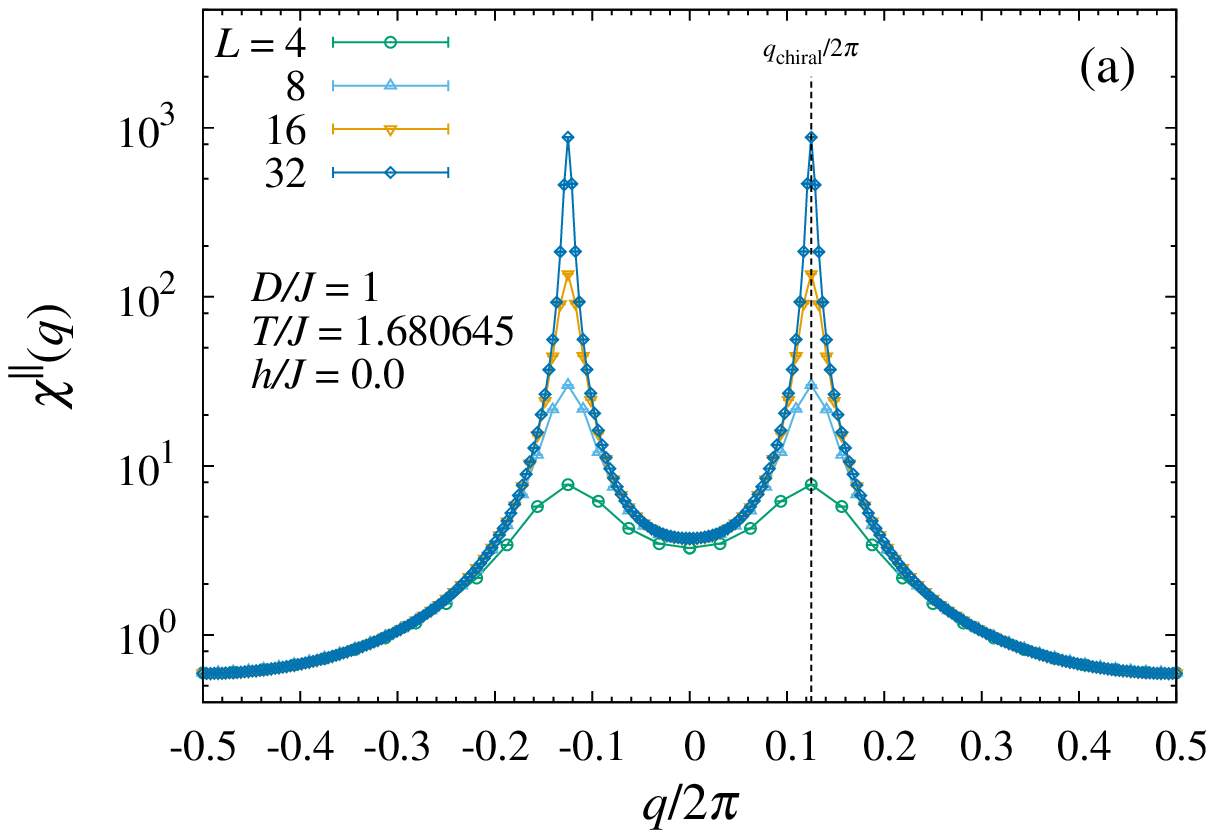}
\includegraphics[width=1.0\linewidth]{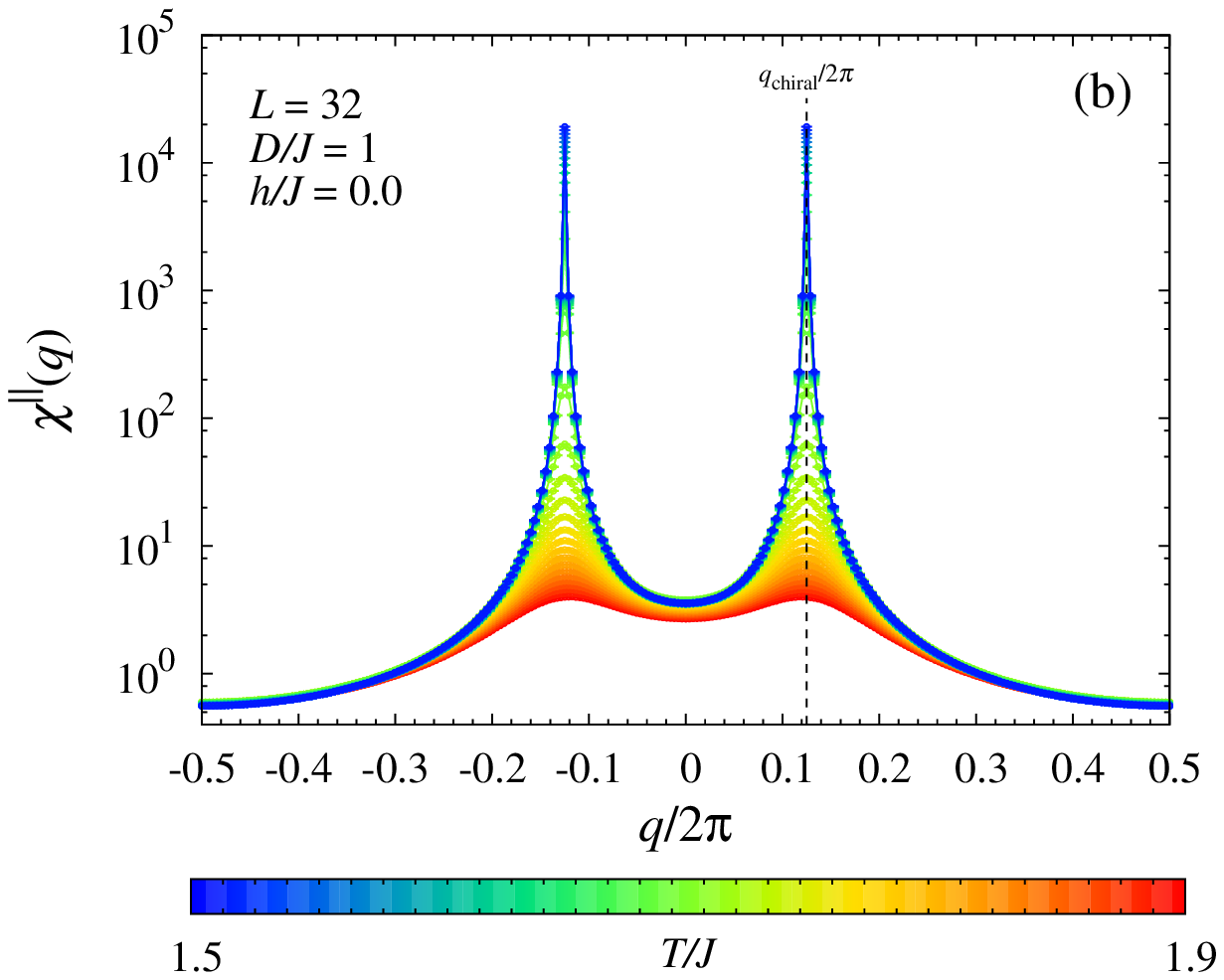}
\caption{(Color online) 
Wave-number dependence of $\chi^{\parallel}\left( q \right)$
of the three-dimensional chiral helimagnetic model without magnetic
 fields 
(a) for various system sizes at $T/J = 1.680645$, which is close to the
 critical temperature,
 and
 (b) with $L = 32$ at various temperatures above and below 
 the critical temperature. 
 }
\label{fig:chi-q_h0}
\end{figure}

\begin{figure}[t]
\includegraphics[width=1.0\linewidth]{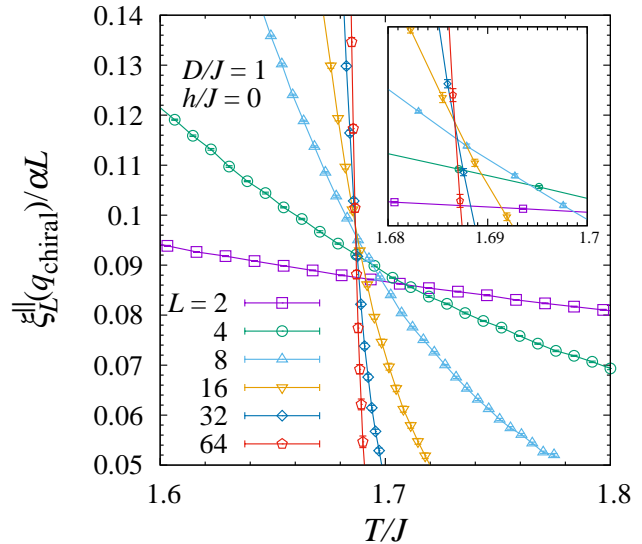}
\caption{(Color online) 
Temperature dependence of 
the finite-size correlation length $\xi_L(q_\mathrm{chiral})$ divided by $\alpha L$
of the three-dimensional chiral helimagnetic model without magnetic
 fields. The inset presents an enlarged view around the critical
 temperature. 
 }
\label{fig:corrlength_h0}
\end{figure}

\begin{figure}[t]
\includegraphics[width=1.0\linewidth]{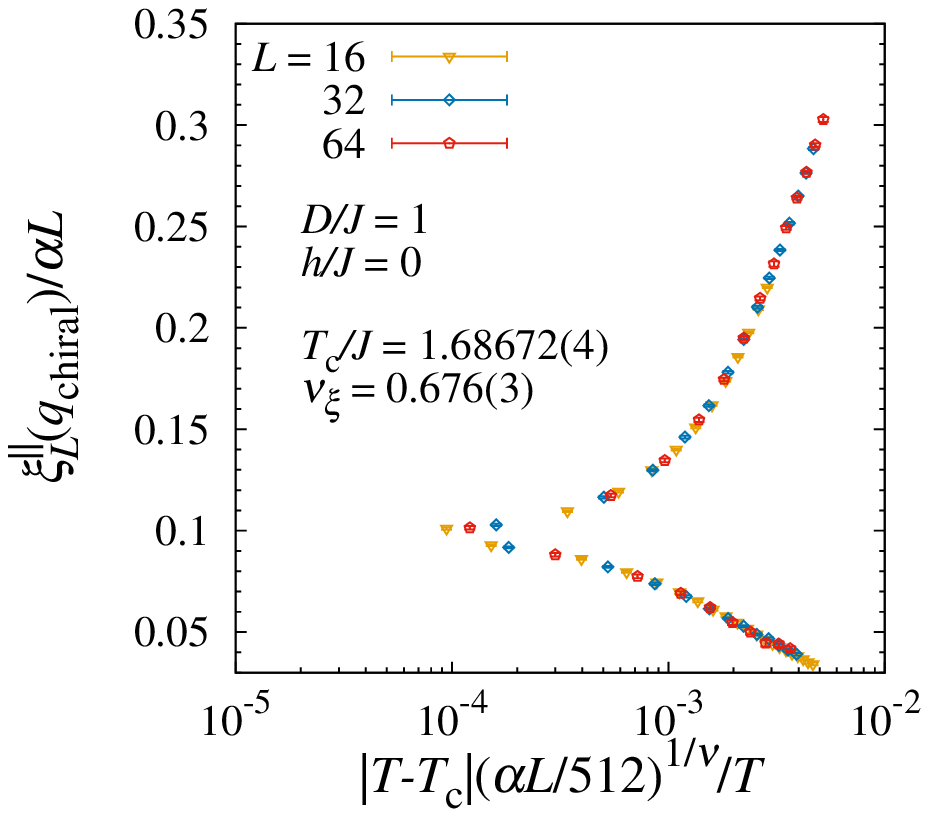}
\caption{(Color online) 
A finite-size scaling plot of the finite-size correlation length 
$\xi^\parallel_L\left( q_\mathrm{chiral}\right)$
divided by $\alpha L$
of the three-dimensional chiral helimagnetic model without magnetic fields.
The smallest system size of this FSS plot is $L_\mathrm{min} = 16$.
The critical temperature $T_\mathrm{c}$ 
and the critical exponent $\nu$ are estimated as 
$T_\mathrm{c} / J = 1.68672(4)$
and $\nu = 0.676(3)$, respectively. 
 }
\label{fig:corrlength-FSS_h0}
\end{figure}

\begin{figure}[t]
\includegraphics[width=0.49\linewidth]{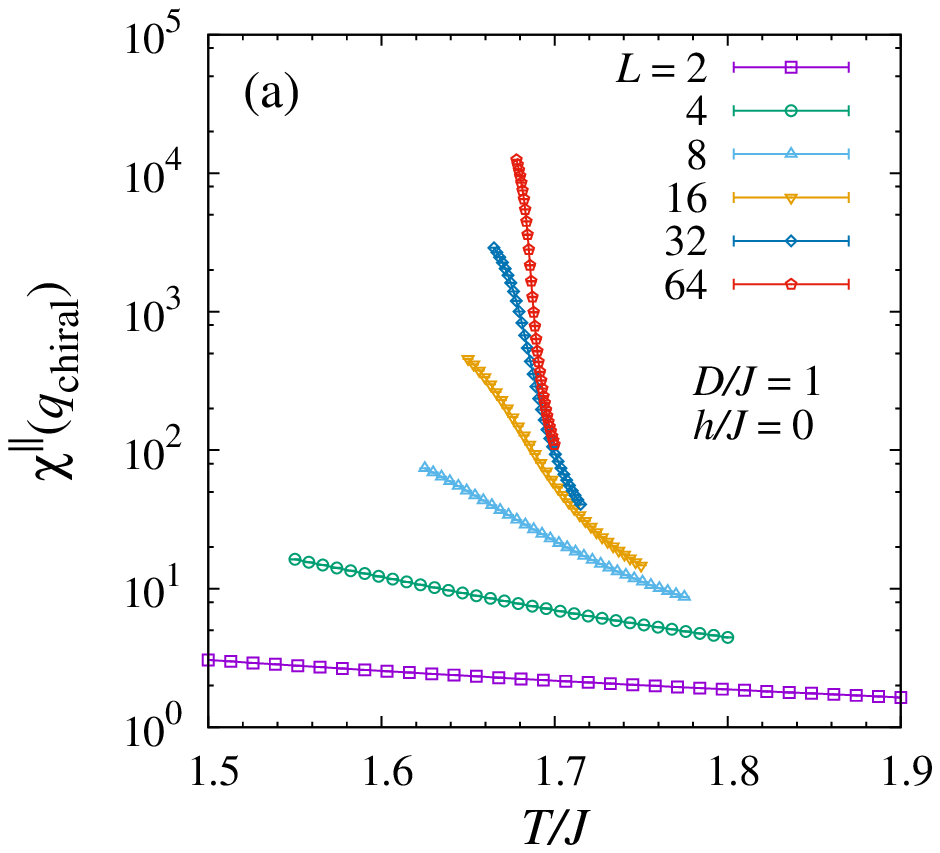}
\includegraphics[width=0.49\linewidth]{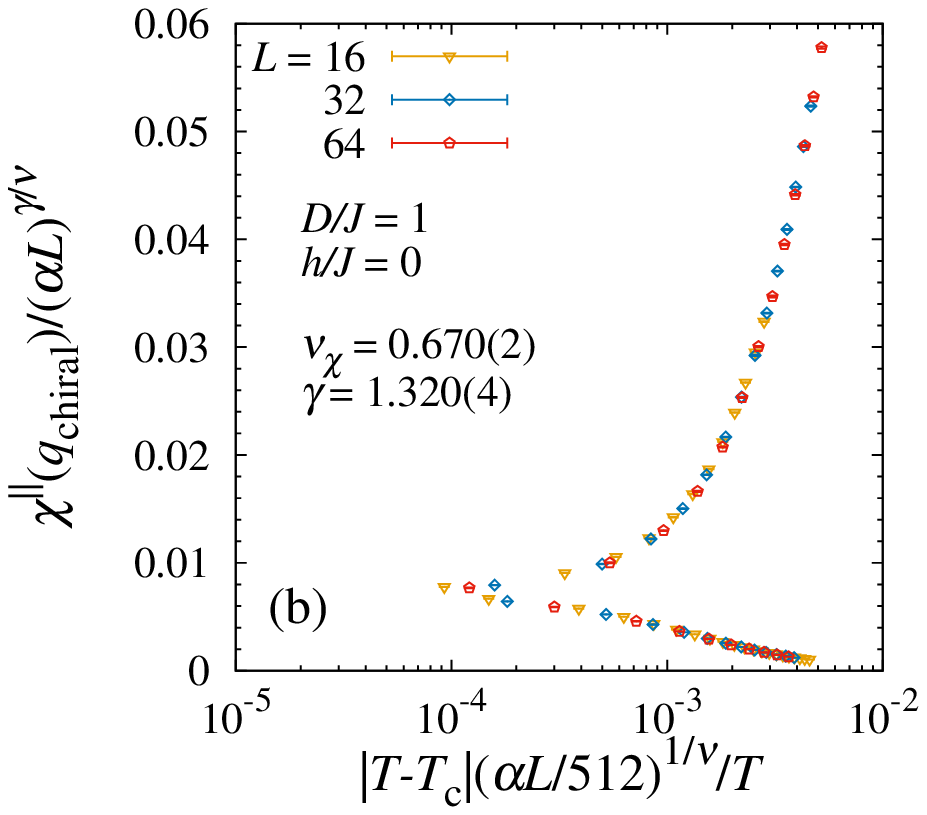}
\caption{(Color online) 
(a):
Temperature dependence of 
the wave-vector-dependent magnetic susceptibility $\chi^\parallel \left(q_\mathrm{chiral}\right)$
of the chiral helimagnetic model in three dimensions without magnetic fields. 
(b):
A finite-size scaling plot of $\chi^\parallel \left(q_\mathrm{chiral}\right)$
of the chiral helimagnetic model in three dimensions without magnetic fields. 
The value of the critical temperature $T_\mathrm{c}$
estimated by the finite-size scaling analysis 
of the finite-size correlation length ratio $\xi^\parallel_L\left( q_\mathrm{chiral}\right) / \alpha L$
is used.
 }
\label{fig:temp-chi-q_h0}
\end{figure}

\section{Result}
\label{sec:result}
In this section, 
we present results of our Monte Carlo simulations of the system
with and without the magnetic field.
The linear size of the system in the simulations
ranges from $L = 2$ (the total number of spins $N = 2 \times 16 \times 2$)
to $L = 64$ ($N = 64 \times 512 \times 64$).
The total number of Monte Carlo steps (MCS)
in our simulations is $5 \times 10^4$ -- $5 \times 10^5$
depending on the system size, 
where one MCS is defined as $N$ lifting events 
with $5$ over-relaxation sweeps per spin.
One heat-bath update per spin is performed for every $10$ MCS.
We checked the equilibration by confirming that
the average values of physical quantities 
measured during an interval
coincide with 
those measured during another interval twice longer
within statistical uncertainty. 
Error bars are evaluated by results of multiple independent simulations.

\subsection{Universality class of the system without magnetic fields}
\label{sec:withoutMag}
First, we present the specific heat $c$ of the system
for various system sizes
in Fig.~\ref{fig:static_h0}.
One can see in the figure
that the specific heat shows a sharp peak
at about $T / J \simeq 1.68$,
and thus,
a phase transition is expected 
to occur at around this temperature.
Around and below this temperature, 
the wave-vector-dependent susceptibility $\chi^\parallel \left( q \right)$
has two peaks at $q = \pm q_\mathrm{chiral}$, see Fig.~\ref{fig:chi-q_h0}.
This fact is insensitive to the system size in our simulations.
Therefore, 
the wave vector $\vec q_\mathrm{chiral}$ also characterizes
the ordering structure of the system at finite temperature
and $\vec m \left( \vec q_\mathrm{chiral}\right)$
can be considered as an order parameter of the system.

We show the wave-vector-dependent finite-size correlation length 
$\xi^\parallel_L \left( q_\mathrm{chiral}\right)$
divided by $\alpha L$ in Fig.~\ref{fig:corrlength_h0}.
One can see in the figure that
each pair of curves for
$\xi_L^\parallel \left( q_\mathrm{chiral} \right) / \alpha L$ 
and $\xi_{2L}^\parallel \left( q_\mathrm{chiral} \right) / 2\alpha L$
intersects at a temperature
and that the intersection converges to a certain temperature point for
larger sizes while it slightly shifts for smaller sizes. 
This implies that 
the correlation length 
with the wave vector $\vec q_\mathrm{chiral}$ diverges
at a finite temperature
in the thermodynamic limit. 
Here, we assume that $\xi_L^\parallel \left( q_\mathrm{chiral} \right) / \alpha L$
follows a finite-size scaling (FSS) form
\begin{equation}
\frac{\xi_L^\parallel \left( q_\mathrm{chiral} \right)}{\alpha L}
= F\left[ \left( T - T_\mathrm{c}\right) \left( \alpha L \right)^{1/\nu}\right],
\end{equation}
where $F$ is a scaling function
and $\nu$ is the critical exponent
of the correlation length.
By using a recently proposed method
based on Bayesian inference \cite{BayesFSS1,BayesFSS2}, 
FSS analyses are performed
for four sets of the data 
consisting of three successive system sizes 
$L_\mathrm{min}$, $2L_\mathrm{min}$ and $4L_\mathrm{min}$.
As shown in Fig.~\ref{fig:corrlength-FSS_h0},
the FSS plot for the data set with $L_\mathrm{min} = 16$ 
works well, yielding that 
the critical temperature $T_\mathrm{c}$ 
and the critical exponent $\nu$ are estimated as
$T_\mathrm{c} / J = 1.68672(4)$
and $\nu = 0.676(3)$, respectively.

Using the value of the critical temperature
estimated by FSS of the finite-size correlation length ratio
$\xi_L^\parallel \left( q_\mathrm{chiral} \right) / \alpha L$,
we also perform FSS analyses
of the wave-vector-dependent susceptibility $\chi^\parallel \left( q_\mathrm{chiral} \right)$
for the same data sets.
The susceptibility is assumed to follow a scaling form
\begin{equation}
\chi^\parallel \left( q_\mathrm{chiral} \right)
= \left( \alpha L \right)^{\gamma/\nu} G\left[ \left( T - T_\mathrm{c}\right) \left( \alpha L \right)^{1/\nu}\right],
\end{equation}
where $G$ is a scaling function
and $\gamma$ is the critical exponent of the susceptibility.
One can see in Fig.~\ref{fig:temp-chi-q_h0} 
temperature dependence of the susceptibility $\chi^\parallel \left( q_\mathrm{chiral} \right)$
and the resultant FSS plot.
The exponents are estimated as
$\nu = 0.670(2)$ and $\gamma = 1.320(4)$,
respectively.
The estimated values of 
the critical temperature and exponents are shown in
Table~\ref{tab:FSS_LTN}.
As seen in the table, 
the values of the critical exponents approach
those of the three-dimensional ferromagnetic \textit{XY} model \cite{3dXY}
as $L_\mathrm{min}$ increases.
We conclude that 
the system without magnetic fields undergoes a phase transition
from a paramagnetic phase
to a chiral helimagnetic phase as temperature decreases
with critical exponents of the three-dimensional \textit{XY} model,
as predicted in Ref.~\onlinecite{RG_CHM1, RG_CHM2, DM-XY}.

\begin{table}[!t]
\begin{tabular*}{1.0\linewidth}{@{\extracolsep{\fill}}ccccc}
\hline
\hline
\multicolumn1{r}{$L_\mathrm{min}$}  
& \multicolumn1{c}{$T_\mathrm{c} / J$} 
& \multicolumn1{c}{$\nu_{\xi}$} 
& \multicolumn1{c}{$\nu_{\chi}$} 
& \multicolumn1{c}{$\gamma$} 
\\ \hline
2 & 1.688(1) &  0.72(2) & 0.711(5) & 1.45(1)\\ 
4 & 1.6871(2) & 0.696(5) & 0.682(2) & 1.314(4) \\ 
8 & 1.68683(5) & 0.681(4) & 0.671(1) & 1.303(3)\\ 
16 & 1.68672(4) & 0.676(3) & 0.670(2) & 1.320(4)\\ \hline\hline
\end{tabular*} 
\caption{
The estimated values of the critical temperature
and the critical exponents of the correlation length
and the susceptibility
by finite-size scaling analyses.
The values of the critical temperature $T_\mathrm{c}$ 
and the exponent of the correlation length denoted as $\nu_\xi$
are estimated using the data 
of the finite-size correlation length $\xi^\parallel_L\left( q_\mathrm{chiral}\right) / \alpha L$.
Using the estimated value of $T_\mathrm{c}$,
the value of critical exponents of the susceptibility $\gamma$ and that
 of the correlation length denoted as $\nu_\chi$ are estimated
by FSS analyses of the susceptibility $\chi^\parallel \left( q_\mathrm{chiral}\right)$.
}
\label{tab:FSS_LTN}
\end{table}

\begin{figure*}[t]
\includegraphics[width=0.32\linewidth]{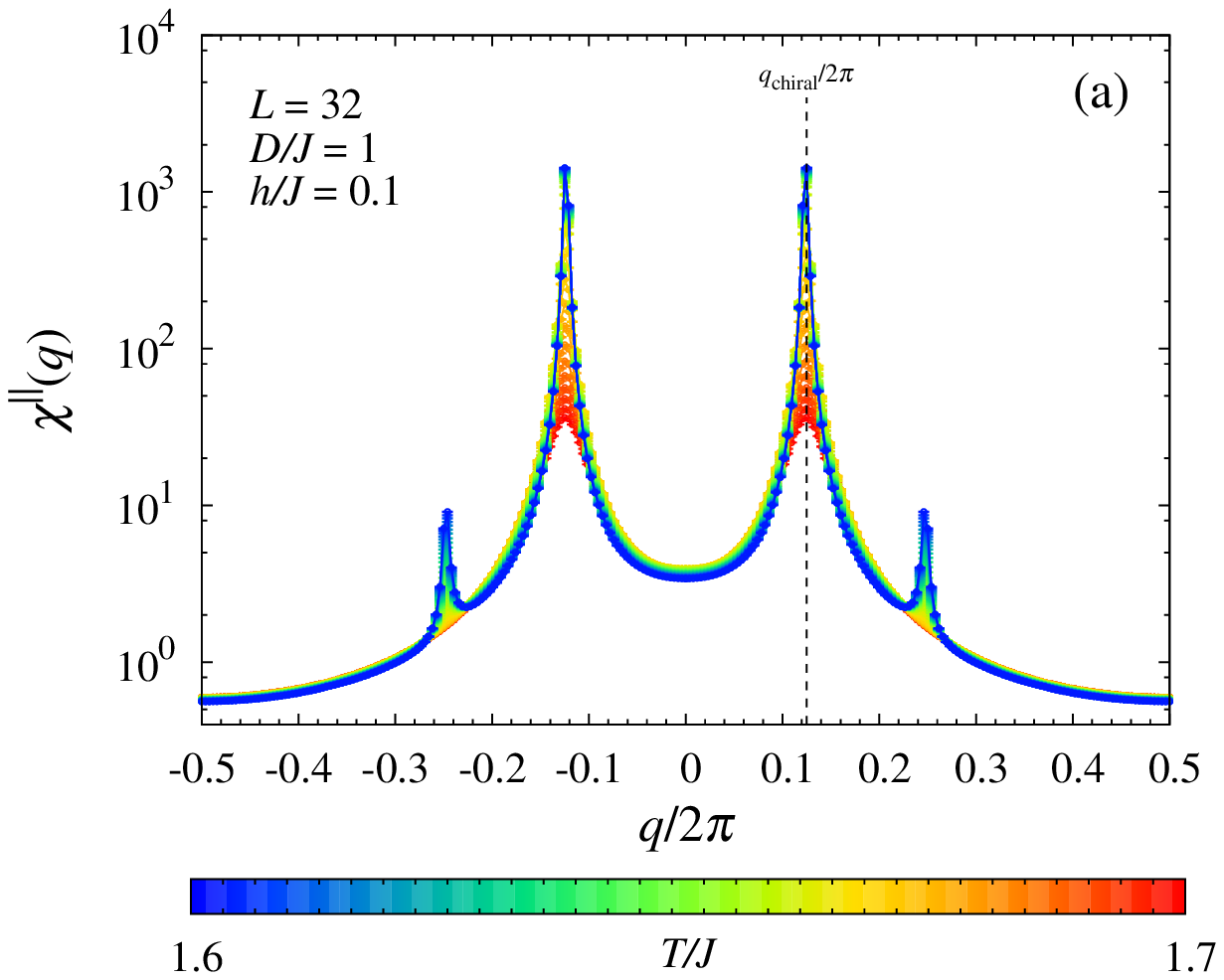}
\includegraphics[width=0.32\linewidth]{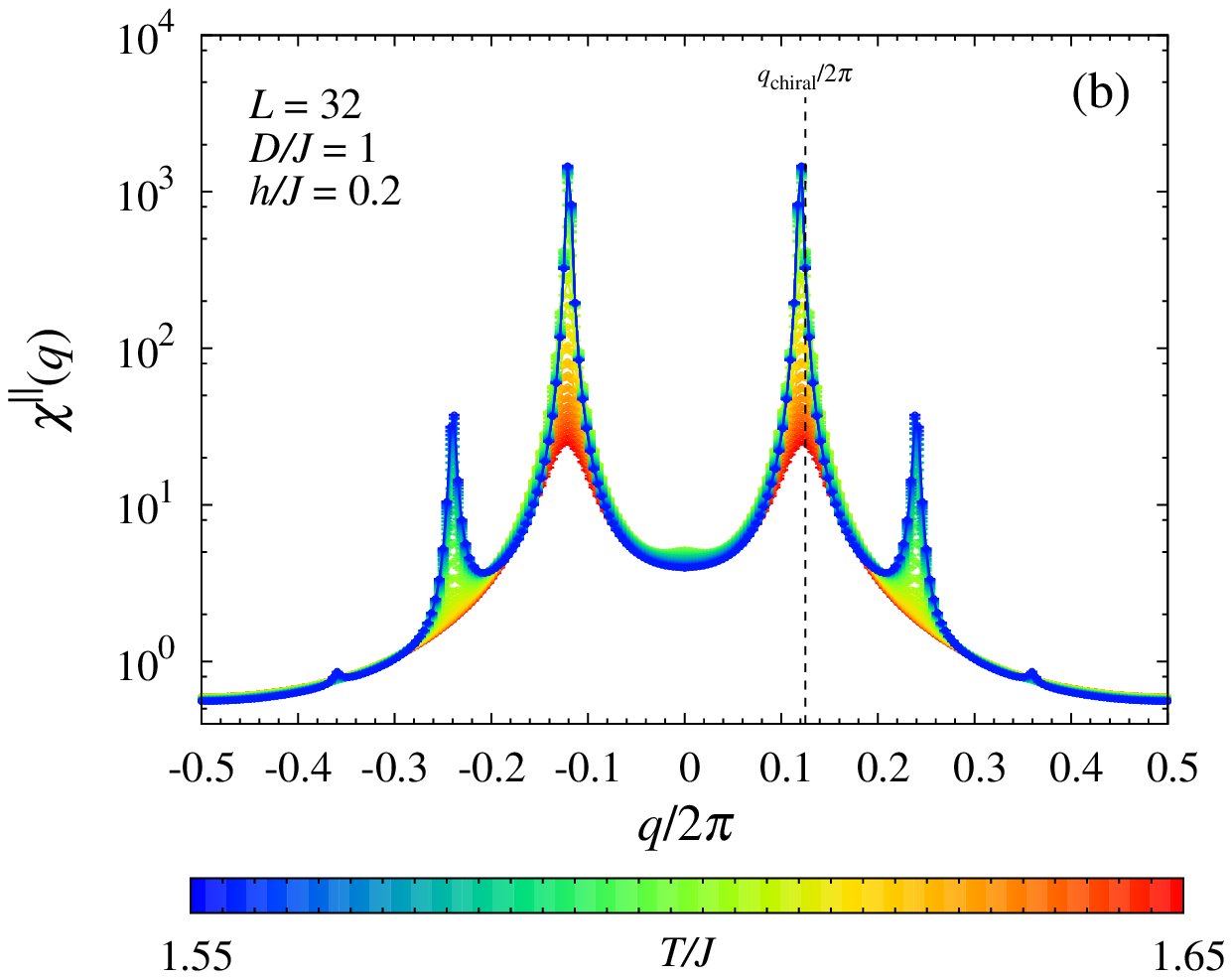}
\includegraphics[width=0.32\linewidth]{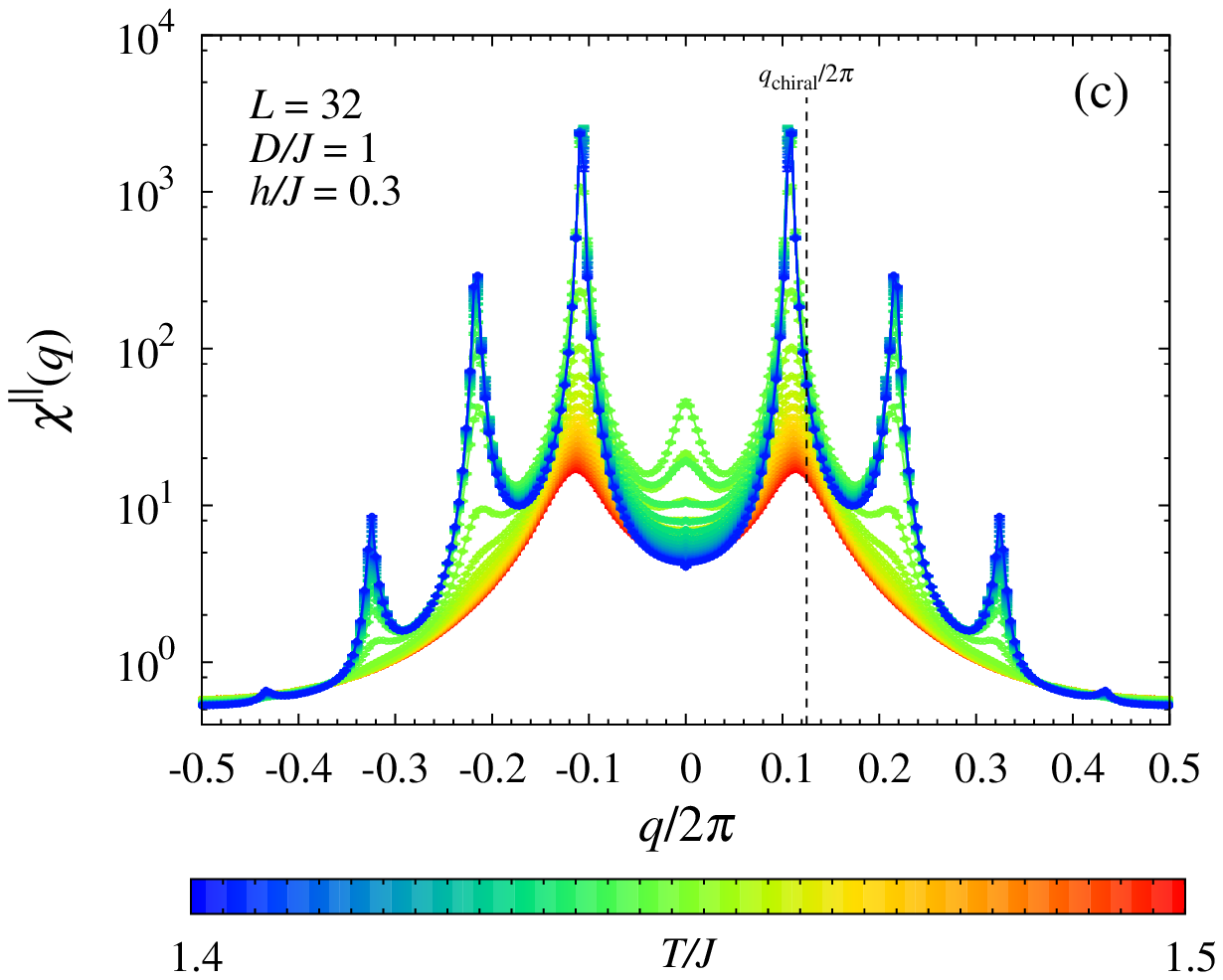}
\caption{(Color online) 
Wave-number dependence of $\chi^\parallel \left( \vec{q} \right)$
of the chiral helimagnetic model in three dimensions
for various temperatures with $L=32$. 
The values of the magnetic fields perpendicular to the DM vector
are (a) $h / J = 0.1$, (b) $h / J = 0.2$, and (c) $h / J = 0.3$.
The vertical line represents $q_\mathrm{chiral}/2\pi$. 
}
\label{fig:chi-q_h0123_tempdep}
\includegraphics[width=0.32\linewidth]{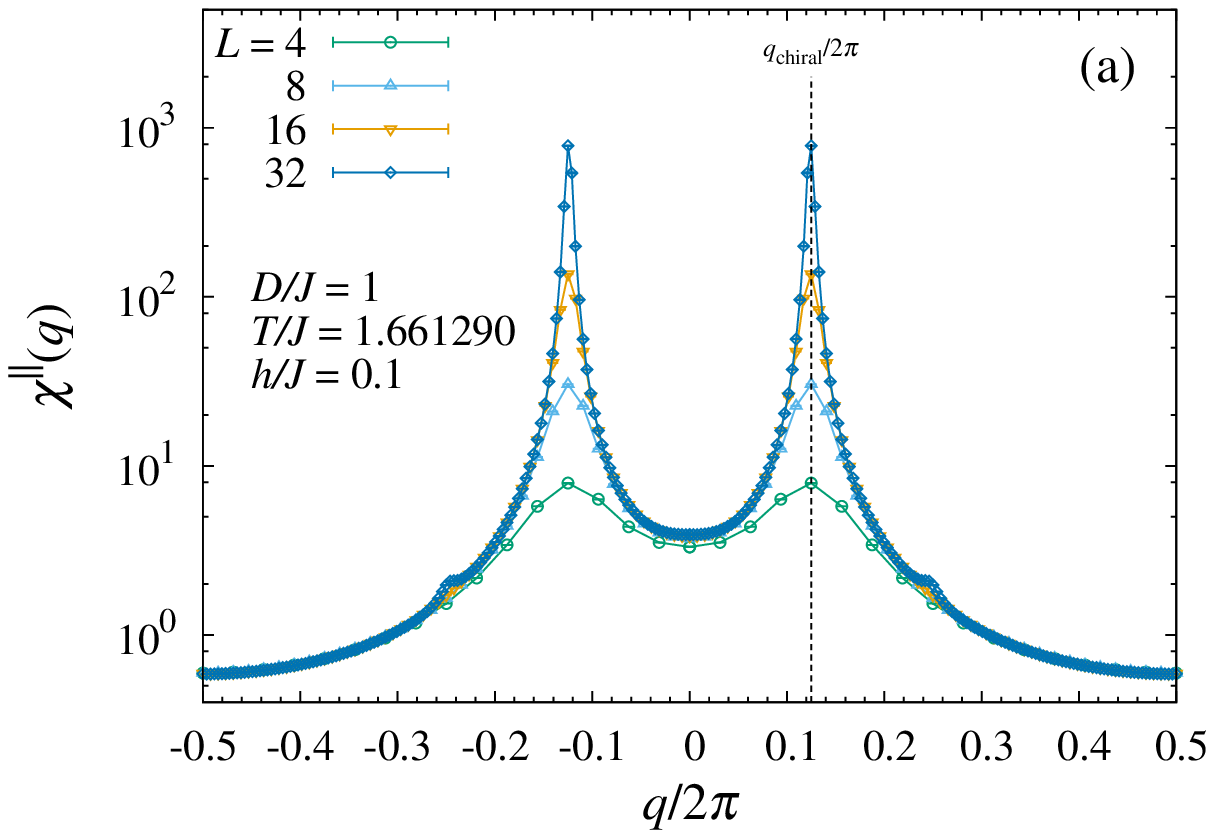}
\includegraphics[width=0.32\linewidth]{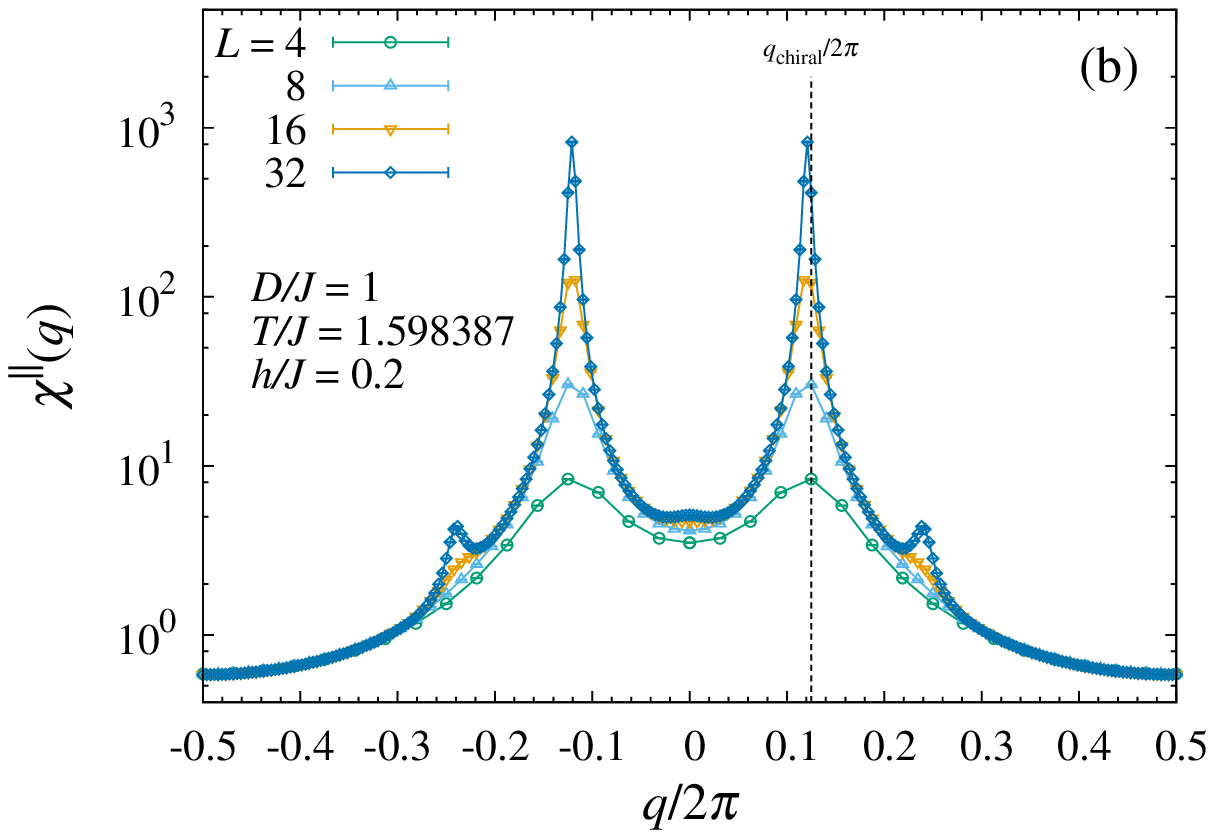}
\includegraphics[width=0.32\linewidth]{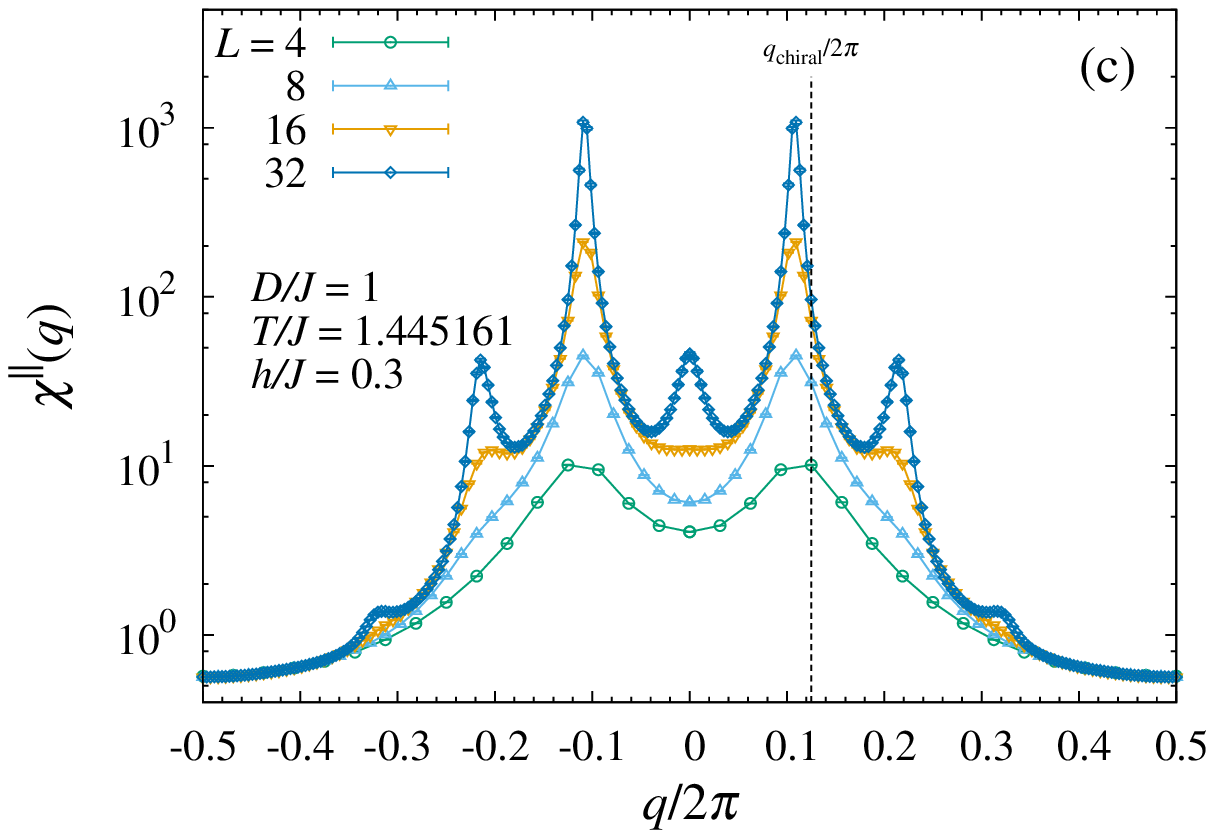}
\caption{(Color online) 
Wave-number dependence of $\chi^\parallel \left( \vec{q} \right)$
of the chiral helimagnetic model in three dimensions
for various system sizes near the estimated transition temperature
depending on the magnetic field. 
The values of the magnetic fields perpendicular to the DM vector
are (a) $h / J = 0.1$, (b) $h / J = 0.2$, and (c) $h / J = 0.3$.
}
\label{fig:chi-q_h0123_sizedep}
\end{figure*}

\begin{figure}[t]
\includegraphics[width=0.8\linewidth]{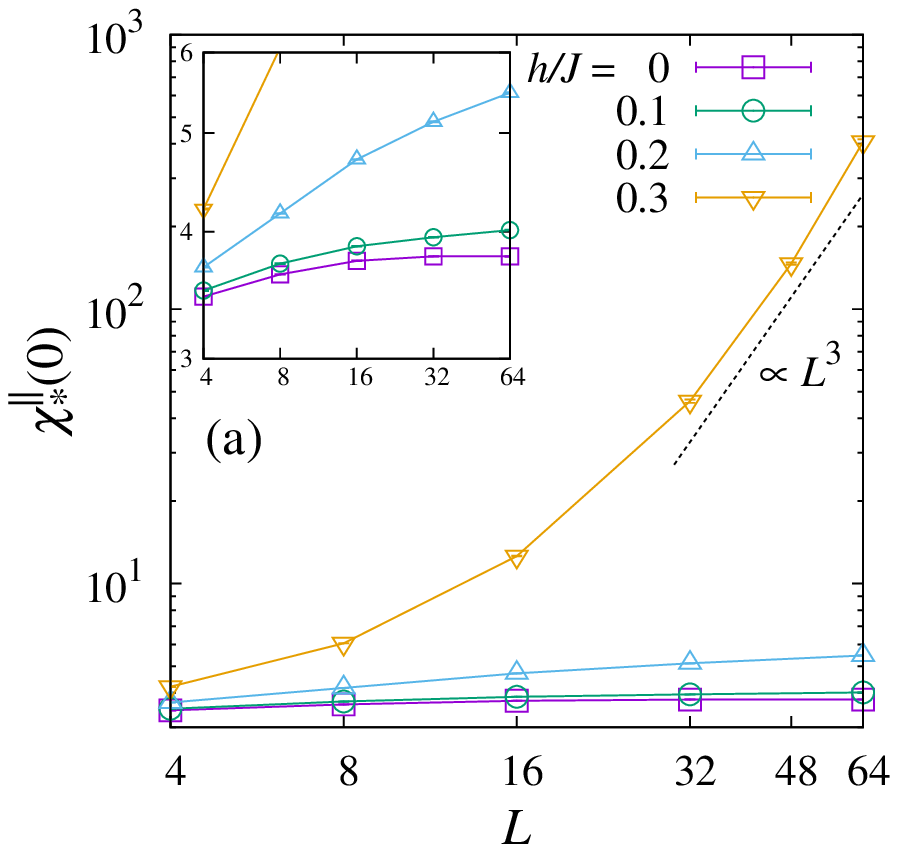}
\includegraphics[width=0.8\linewidth]{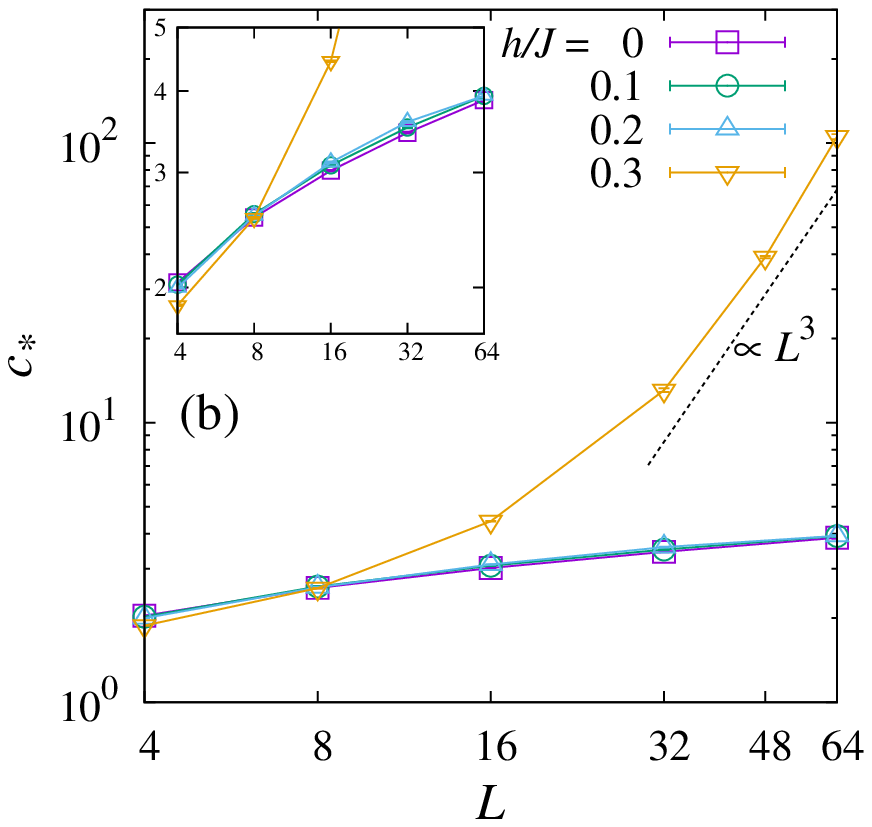}
\caption{(Color online) 
System-size dependence 
of the peak value of the susceptibility $\chi^\parallel_* \left( 0\right)$ (a)
and the specific heat $c_*$ (b)
of the chiral helimagnetic model in three dimensions
with a magnetic field perpendicular to the DM vector 
$h / J = 0$, $0.1$, $0.2$, and $0.3$.
The black dotted lines are proportional to $L^3$.
The insets show enlarged views.
 }
\label{fig:chi0_size-dep}
\end{figure}

\begin{figure}[t]
\includegraphics[width=0.8\linewidth]{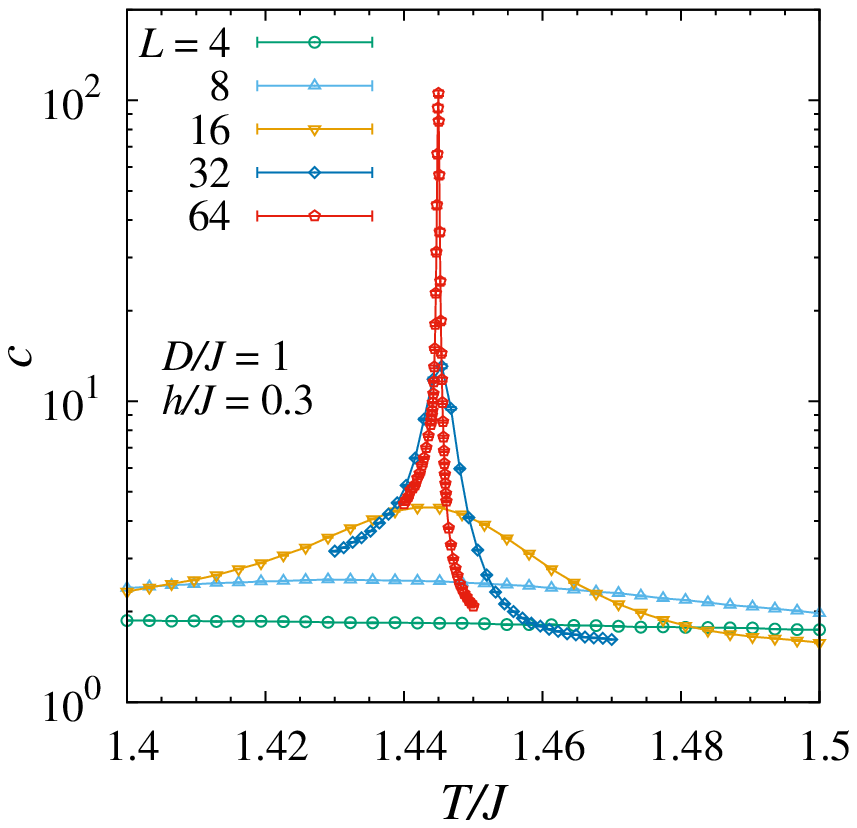}
\caption{(Color online) 
Temperature dependence of 
specific heat $c$ 
of the chiral helimagnetic model in three dimensions
with a magnetic field perpendicular to the DM vector $h / J = 0.3$.
 }
\label{fig:static_h03}
\includegraphics[width=1.0\linewidth]{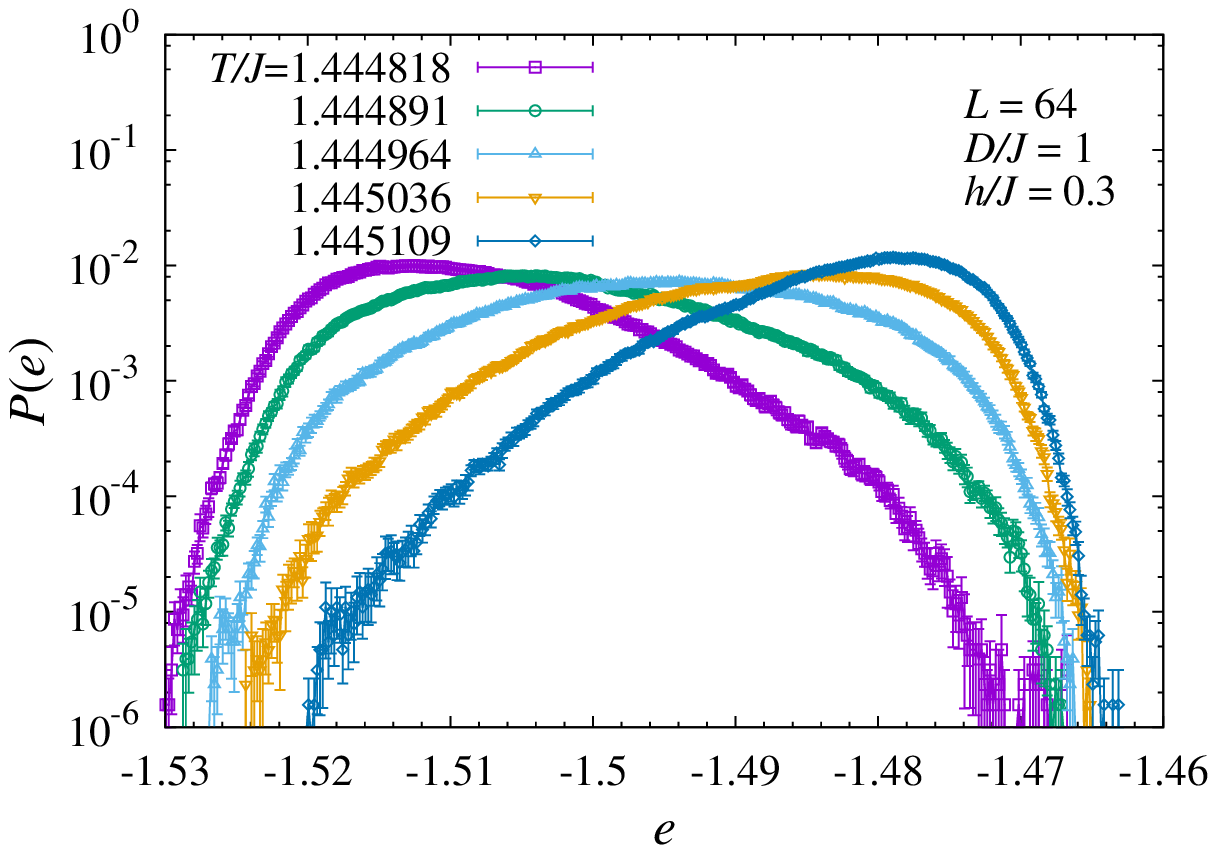}
\caption{(Color online) 
The energy-density distribution function $P\left( e\right)$
of the chiral helimagnetic model in three dimensions
with a magnetic field perpendicular to the DM vector $h / J = 0.3$.
The system size $L = 64$ is the largest size in our simulations and the temperatures are close to the transition temperature.
 }
\label{fig:static_P_e-h03}
\end{figure}

\subsection{Phase transition under a magnetic field perpendicular to the DM vector}
\label{sec:withMag}
In this subsection,
we focus on the effect of a magnetic field perpendicular to the DM
vector. 
The wave-number dependence
of the susceptibility $\chi^\parallel \left( q \right)$
at $h / J = 0.1$, $0.2$, and $0.3$  
for various temperatures and various sizes is shown
in Fig.~\ref{fig:chi-q_h0123_tempdep} 
and Fig.~\ref{fig:chi-q_h0123_sizedep},
respectively.
In contrast to the case without the magnetic field shown in
Fig.~\ref{fig:chi-q_h0}, 
$\chi^\parallel \left( q \right)$ has several peaks 
at $\pm q_0$ and integral multiples of $q_0$
in the presence of the magnetic field
in the low temperature region 
with $q_0$ being the positive wave number 
which gives the largest value of the susceptibility. 
The value of $q_0$ for finite magnetic fields is significantly smaller
than that of $q_\mathrm{chiral}$, although the difference is tiny for
small fields as shown in Fig.~\ref{fig:chi-q_h0123_tempdep}  and
Fig.~\ref{fig:chi-q_h0123_sizedep}. 				 
Furthermore,
not only the largest peaks
but also other small peaks
are enhanced 
with increasing the system size,
as seen in Fig.~\ref{fig:chi-q_h0123_sizedep}.
These indicate that
a periodic order, e.g., chiral soliton lattice (CSL)
which cannot be characterized by a single wave vector
emerges at low temperatures  
in the thermodynamic limit.
The distance 
between two chiral solitons
in the low temperature region
is characterized by 
the value of the wave number $q_0$
as $\sim 2 \pi / \left| q_0 \right|$.
In Fig.~\ref{fig:chi-q_h0123_tempdep}(c),
for instance,
one can see
that $\left| q_0\right| / 2 \pi \sim 0.1$ at a sufficiently low temperature
for $h / J = 0.3$,
and hence,
the distance between two chiral solitons along the DM vector
is about $10$ lattice spacings. 
Other wave numbers 
of the peak in $\chi^\parallel \left( q\right)$
in the low temperature region
are considered to characterize
shorter length scales 
within one chiral soliton.

One may consider naively the order parameter of the CSL order to be
$\vec{m} \left( \vec q_0 \right)$.
The value of $q_0$ weakly depends on temperature
and also  the values of the wave numbers of the peaks
in finite systems with the magnetic field
slightly deviate from those in the thermodynamic limit.
The latter is due to the fact that 
the wave number in finite-size lattices
can take only discrete values.
As discussed above, the existence of the CSL phase
characterized by the multiple wave vectors is strongly suggested at low
temperatures.
It is, however, difficult to identify the precise value of $q_0$ in
numerical simulations and the order parameter in the CSL phase.

While the CSL emerges
in the presence of the magnetic field,
qualitatively different behavior is observed in thermodynamic quantities
at a relatively large
magnetic field, particularly at $h / J = 0.3$ in our study. 
One of the striking features is the existence of the sharp peak of
$\chi^\parallel\left(0 \right)$ at a certain temperature  
which is not the intrinsic susceptibility conjugated with the CSL
order and
also the chiral helimagnetic order parameter.
At the temperature, the specific heat has a diverging peak
simultaneously. 
We show in Fig.~\ref{fig:chi0_size-dep}
the system-size dependence 
of the peak values
of the magnetic susceptibility $\chi^\parallel_* \left( 0 \right)$
and the specific heat $c_*$.
For $h / J = 0.1$ and $0.2$,
the peak values of $\chi^\parallel_* \left( 0 \right)$ and $c_*$
do not seem to diverge 
even in the thermodynamic limit.
This is compatible with the result of 
$h / J = 0$,
where the system belongs to the universality class
of the three-dimensional \textit{XY} model and hence the critical
exponent $\alpha$ is negative. 
Without the magnetic field, 
the specific heat $c$ does not diverge,
but shows a cusp singularity at the critical  temperature 
in the thermodynamic limit 
as the three-dimensional \textit{XY} model.
When a cusp singularity exists in the specific heat, 
its peak value $c_*$ scales as \cite{PeczakFerrenbergLandau, HolmJanke}
\begin{equation}
c_* \simeq c_*^\infty - s L^{\alpha / \nu},
\end{equation}
where $c_*^\infty$ is the peak value of the specific heat in the thermodynamic limit
and $s$ is a constant. 
We can see in the inset of Fig.~\ref{fig:chi0_size-dep}(b) 
that the peak values $c_*$ of the system with $h / J = 0$, $0.1$ and $0.2$
have very similar system size dependence.
This fact suggests that the system under the magnetic fields 
also belongs to the universality class
of the three-dimensional ferromagnetic \textit{XY} model.

On the other hand,
for $h / J = 0.3$,
the peak values $\chi^\parallel_* \left( 0 \right)$ and $c_*$
show very strong tendencies to diverge
in the thermodynamic limit.
In particular, 
$\chi^\parallel_* \left( 0\right)$ and $c_*$ at $h / J = 0.3$
seem to diverge as a power law with $L^3$ or even faster than a power
low in larger system sizes.
These indicate the existence
of a critical point $\left(T_\mathrm{d}, h_\mathrm{d} \right)$
where $0.2 < h_\mathrm{d} / J < 0.3$
on the phase boundary
between the paramagnetic phase and the CSL phase
in the magnetic phase diagram
of the system.
In other words, 
the system is expected to
have finite values
of the specific heat $c$ and
the susceptibility $\chi^\parallel \left( 0\right)$
at the transition temperature 
for $h < h_\mathrm{d}$, 
and presumably belongs to the same universality class
of the system without the magnetic field,
while the system undergoes a phase transition
at a finite temperature
with the diverging specific heat $c$
and diverging magnetic susceptibility $\chi^\parallel \left( 0 \right)$
for $h > h_\mathrm{d}$. 

A possible explanation of the strong divergence of the specific heat
found at $h / J = 0.3$ might be an occurrence of the
first-order phase transition. 
Then, the specific heat has a delta-function type divergence at the
transition temperature and 
the peak value of the specific heat is expected to diverge  as $L^d$ 
where $d = 3$ is the spatial dimension \cite{first-order}. 
Also the energy-density distribution has two peaks at the transition
temperature. 
In Fig.~\ref{fig:static_h03},
we present
temperature dependence of 
the specific heat $c$
of the system with $h / J = 0.3$. 
One can see in the phase diagram that
the specific heat $c$ shows 
a very sharp peak
at about $T / J \simeq 1.445$,
and the width of the peak
becomes narrower
as the system size increases.
This is consistent with the occurrence of the first-order transition and
the size dependence of $c_*$ shown in Fig.~\ref{fig:chi0_size-dep}(b) is
marginally compatible with $L^3$. 
However, 
as seen in Fig.~\ref{fig:static_P_e-h03},
the energy-density distribution function $P \left( e \right)$
does not have a double-peak structure
near the transition temperature.
No clear evidence of the first-order transition is found in our
numerical results.
We could not completely rule out the possibility of a weak first-order
transition with a finite correlation length at the transition temperature
larger than the largest system size in our simulations. 
Therefore, we tentatively conclude that
this phase transition found at $h/J=0.3$ 
is a continuous one. 
Our results suggest that 
the expected universality class has a ratio of the critical exponents of
the specific heat and the correlation length  $\alpha / \nu > 3$, 
assuming that $c_*$ of the system diverges faster than $L^3$ also in
larger systems.
Unfortunately, we could not 
determine the critical exponents of the transition
and the precise location of the critical point
$\left( T_\mathrm d, h_\mathrm d \right)$, which 
requires larger scale simulations of the system.

\section{Discussion and Summary}
\label{sec:discussion}

\begin{figure}[]
\includegraphics[width=1.0\linewidth]{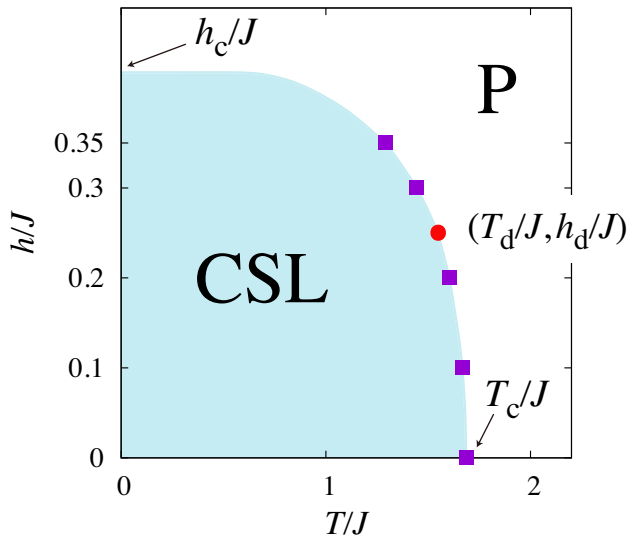}
\caption{(Color online) 
A possible magnetic phase diagram
of the chiral helimagnetic model in three dimensions.
In the phase diagram, ``CSL'' and ``P'' denote the chiral soliton
lattice phase and paramagnetic phase, respectively.
The filled squares are estimated transition temperature in this  work
and the circle represents an expected critical point whose precise
location is not determined. 
}
\label{fig:phase-diag}
\end{figure}

A possible phase diagram of the system
is presented in Fig.~\ref{fig:phase-diag}, 
where 
we denote the paramagnetic phase and the CSL phase
as ``P'' and ``CSL'', respectively.
The filled square at $h / J = 0$
is estimated by the FSS analysis
in Sec.~\ref{sec:withoutMag},
and other squares are 
estimated by the peak temperature of 
$\chi^\parallel \left( 0 \right)$
at $h / J = 0.1$, $0.2$ and $0.3$ for $L = 64$ and 
at $h / J = 0.35$ for $L = 16$.
The circle represents 
an expected 
location of
the critical point $\left( T_\mathrm{d} / J, h_\mathrm{d} / J \right)$. 

One can see in the phase diagram that 
the phase boundary $h_{\partial \mathrm{CSL}} \left(T \right)$ 
between the paramagnetic phase and the CSL phase 
has a finite slope, 
which is compatible with 
the experimental phase diagram of a chiral helimagnet \cite{CSL_CHM_ex2}.
Imposing differentiability on the free-energy density
of the infinite system
at a point $\left( T_0, h_{\partial \mathrm{CSL}} \left( T_0 \right)\right)$ 
where a second-order phase transition occurs,
the finite tangent of the phase boundary yields the relation 
\begin{equation}
\Delta \chi \Delta c - T \left( \Delta \omega \right)^2 = 0, 
\label{eq:phaseb}
\end{equation}
where 
$\omega$ and $\chi$
are the temperature derivative and the magnetic-field derivative
of the magnetization parallel to the field, 
and $\Delta X = X_{\rm CLS} - X_{\rm P}$ for any $X \in \{ c, \chi, \omega \}$ 
at $\left( T_0, h_{\partial \mathrm{CSL}}\left( T_0\right)\right)$, respectively.
If the system under the magnetic field 
with $0 < h < h_\mathrm d$ belongs to the universality class
of the three-dimensional ferromagnetic \textit{XY} model
as discussed above,
the specific heat is continuous on the phase boundary.
In this system for a fixed $h < h_\mathrm d$, 
the uniform susceptibility has a finite value.
Therefore, Eq.~(\ref{eq:phaseb}) requires $\Delta \omega = 0$,
meaning that the magnetization parallel to the magnetic field
is smooth at the transition temperature.

For $h > h_{\mathrm d}$, however,  
the strong divergence is found in the specific heat.
The difference $\Delta c$ is infinitely large
unless the critical amplitude ratio is
accidentally 1 with the same critical exponent above and below the
critical temperature which may unlikely occur in finite dimensions.
Then, the relation of Eq.~(\ref{eq:phaseb}) allows typically two cases: (i)
$\Delta \chi = 0$ and $\Delta \omega$ is finite and (ii)
$\Delta \chi = \infty$ and $\Delta \omega=\infty$.
Our result of the divergence of $\chi^\parallel \left( 0 \right)$
indicates the latter case.
Precisely speaking, $\chi$ is not identical with 
$\chi^\parallel \left( 0 \right)$ but $\Delta \chi$ 
likely diverges when $\chi^\parallel \left( 0 \right) = \infty$. 
This implies that the exponent of the divergence of 
$\chi^\parallel \left( 0 \right)$ coincides with that of the specific heat. 
Furthermore, the temperature dependence of the magnetization
is also described by the same singularity at least either above or below 
the critical temperature.
Thus, the critical singularity of the specific heat appears in other
observables unrelated to the critical nature through the relation of
Eq.~(\ref{eq:phaseb}), 
while in a conventional system where $\chi$ is an order-parameter
susceptibility, the relation yields the scaling relation
$\alpha+2\beta+\gamma=2$ among the critical indices.

We should note here that
Dzyaloshinskii predicts 
by analyzing the one-dimensional continuum model of the chiral helimagnet
in the presence of the magnetic field
that a continuous phase transition
occurs at a finite temperature \cite{IEDzyalo65}.
It is also shown 
that the specific heat diverges from below the transition temperature
with a logarithmic correction as
\begin{equation}
c \propto \frac{1}{\left(T_* - T \right) \log^2 \left( T_* - T\right)},
\end{equation}
where $T_*$ is the transition temperature,
while no divergence of $c$ displays from above $T_*$.
In this case, $\Delta c$ is infinity at $T_*$
and the critical exponent of the specific heat $\alpha^\prime = 1$ below $T_*$ and
$\alpha = 0$ above $T_*$.
Although no definite conclusion can be drawn on the validity of this
peculiar prediction, our numerical data of the specific heat is not
inconsistent with the asymmetric behavior between above and below the
critical temperature.
One of the main difficulties 
in determining the critical indices is due to the logarithmic-correction term,
which makes the critical region narrow. 
Assuming 
the hyperscaling relation 
$d \nu = 2 - \alpha$
and $\alpha=1$, 
the critical exponent of the correlation length is $\nu = 1 / 3$,
and hence,
the peak value of the specific heat 
is  expected to diverge as $\sim L^{\alpha/\nu} = L^3$.
It also coincides with that in the system with the first-order transition.
As discussed in \ref{sec:withMag}, the power-law divergence of $c_*$
with $L^3$ is marginally consistent with our numerical result. 
Further investigations are required to clarify
the nature of the phase transition
of the system with $h > h_\mathrm d$
and examine the validity of 
Dzyaloshinskii's theory \cite{IEDzyalo65}.

In summary,
we have numerically studied
the classical Heisenberg spin model of a chiral helimagnet
in three dimensions
by equilibrium Monte Carlo simulations 
using the event-chain algorithm.
We have particularly focused on 
its finite-temperature phase transitions with and without
a magnetic field perpendicular 
to the axis of the helical structure.
Without the magnetic field,
it is shown by the FSS analysis that
the system undergoes a continuous phase transition
with critical exponents 
of the three-dimensional ferromagnetic \textit{XY} model
as predicted by some theoretical studies.
It is found that the nature of phase transitions
changes in the presence of the magnetic field,  
although we speculate that 
the phase transition 
is continuous
irrespectively with 
the value of the magnetic field $h$.
While the specific heat $c$ and the magnetic susceptibility $\chi^\parallel \left( 0\right)$
have finite values
at the transition temperature
for  $h / J = 0.1$ and $0.2$, 
they diverge at the transition temperature
for $h / J = 0.3$.
Consequently, it is 
suggested that 
the critical point $\left( T_\mathrm{d}, h_\mathrm{d}\right)$
exists in the region where $0.2 < h_\mathrm{d} / J < 0.3$ in the phase
diagram of the system. 
The critical exponents of the phase transitions
at and above $h_\mathrm d$
remain unclear,
and thus it would be 
interesting
to reveal the universality class of the phase transition in high
fields by determining the critical exponents.
A promising way for studying the phase structure might be the method of
renormalization group. 
Our results suggest that the phase transition, distinct from the
transition at the low fields, can be detected as a
strong singularity in the specific heat, uniform susceptibility and also
magnetization curve, which are measurable in experiments.  
However, the amplitude of the DM interaction studied in this paper is rather large
from viewpoint of experiments. Thus, the dependence of the critical
point is to be clarified in comparison with the experiments.

\acknowledgements{
The authors thank S.~Hoshino and Y.~Kato for very useful discussions
and S.~Takabe for carefully reading the manuscript.
Numerical simulation in this work has mainly been
performed by using the facility of the Supercomputer Center, Institute
for Solid State Physics, the University of Tokyo. 
This research was supported by the Grants-in-Aid for Scientific Research 
from the {JSPS}, 
Japan (No. 25120010  and 25610102), and 
JSPS Core-to-Core program ``Nonequilibrium dynamics of soft matter and information.''
This work was also supported by “Materials research by Information Integration” Initiative (MI$^2$I) project of the Support Program for Starting Up Innovation Hub from Japan Science and Technology Agency (JST).
}

\nocite{*}
\bibliography{ECMC2nd_ref}

\end{document}